# A Universal Vortex-Formation Law for Insect Hovering

David Greenblatt

Technion – Israel Institute of Technology, Haifa, 3200003, Israel

Flapping insect wings generate leading-edge vortices (LEVs) [1] that produce much of the lift required for hovering, yet the frequency governing their formation has not been connected to classical vortex shedding. Roshko's scaling for bluff-body wakes [2], together with Sigurdson's extension to wall-bounded separated flows [3], identifies a universal vortex-formation Strouhal-number range of approximately 0.15–0.17. Here we show that this same range governs LEV formation on both stationary wings and hovering insect wings. We first develop an approximate scaling for stationary flat-plate wings in which discrete LEVs are generated by periodic leading-edge perturbations [4], and show that maximum lift consistently occurs within the universal Strouhal range. In hovering insects, one LEV forms during each translational half-stroke, allowing the same framework to be applied to published flight data. Despite large differences in morphology, kinematics and Reynolds number, the insect data recover the same Strouhal range. These results reveal a common vortex-formation timescale linking classical separated flows, artificially generated LEVs and the natural LEVs that sustain insect hovering.

The repeated formation of alternating-sign leading-edge vortices (LEVs) is a defining feature of hovering insect flight [1]. Their role in generating high lift has been established by experiments, computational studies and dynamically scaled models [5-10]. Navier-Stokes simulations of a hovering bumblebee with flapping and revolving wings illustrate this mechanism directly (Figure 1): the LEV identified by the $Q$-criterion above the wing is associated with a strong region of low pressure over the upper surface, thereby contributing substantially to the lift required for hovering [11]. Although LEV formation and stability have been examined across a broad range of insects and Reynolds numbers, the characteristic frequency of this vortex-generation process lacks a physics-based dimensionless parameter. Specifically, the biomechanics Strouhal number $f_f A/\bar{V}$ [12], useful for correlating cruising animal flight, is of little use because the numerator and denominator simply cancel: here $f_f$ is the wingbeat frequency, $A$ is the peak-to-peak flapping amplitude, and $\bar{V}=2f_f\Phi R$, where $\Phi$ and $R$ are the wingtip flapping angle and semispan, respectively [8]. An alternative parameter is the well-known reduced frequency $k_f \equiv \pi f_f c_m / \bar{V}$, where $c_m$ is the mean chord-length. This represents the ratio of oscillatory and convective timescales, but it is simply borrowed from classical aerodynamics and conventional dynamic stall studies [13,14]. Furthermore, upon substitution, $k_f = \pi / 2\Phi\Lambda_s$, where $\Lambda_s \equiv c_m / R$ is the semispan aspect ratio; thus it characterizes only wing geometry and kinematics, with no direct relationship to LEV generation [8].

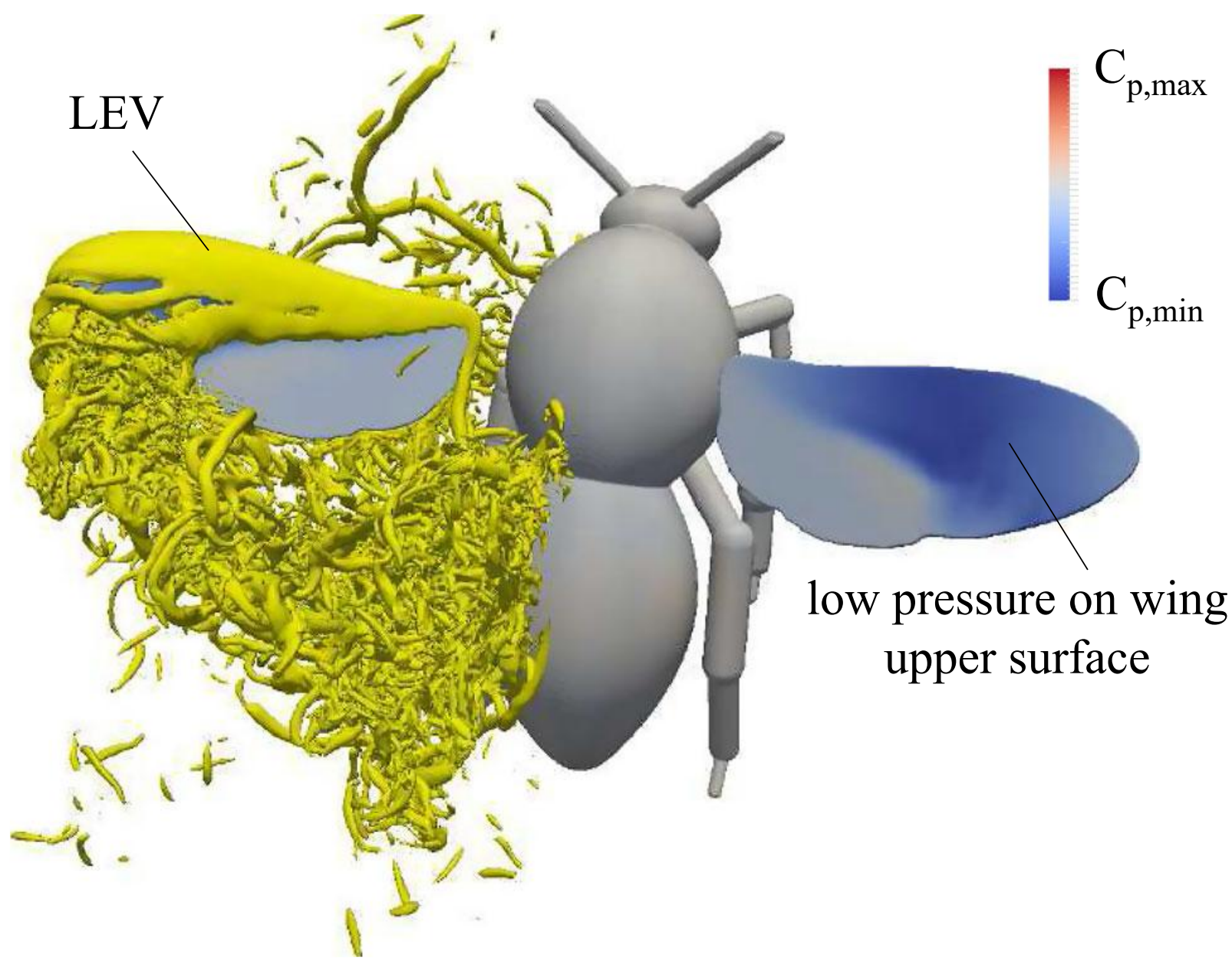

**Figure 1 | Leading-edge vortex and aerodynamic loading of a hovering bumblebee.** Navier-Stokes simulation of a hovering bumblebee with flapping and revolving wings from Liu et al. [11]. Left-wing: vortical structures identified using a $Q$-criterion isosurface, showing the prominent leading-edge vortex (LEV) above the wing. Right-wing: pressure-coefficient distribution on the opposite wing, showing the low-pressure region associated with LEV-induced aerodynamic loading. Republished with permission.

Our approach follows from the vortex-shedding scaling of separated bluff-body flows. Roshko [2] showed that vortex shedding from two-dimensional bluff bodies of different geometries, with a primary dimension $D$, occurs over a narrow range when the shedding frequency is scaled using the cross-wake dimension $D'$ and the velocity at separation $U_{sep}$. His experiments spanned approximately $400 \leq Re \equiv U_\infty D / \nu \leq 20{,}000$, notably encompassing the hovering Reynolds numbers of a broad range of insects. Sigurdson [3] subsequently extended this universal scaling to wall-bounded separated and reattaching flows. Here, the natural shedding frequencies exhibited similar scaling with $D' = 2h_b$, where $h_b$ is separation-bubble height and can also be interpreted as the characteristic vortex dimension. Thus, a common vortex-frequency scaling emerged based on the cross-wake dimension and the convection velocity of the separated region, namely

$$0.15 \le St \equiv \frac{f_{\text{vortex}} D'}{U_{\text{sep}}} \le 0.17,$$

where $f_{\text{vortex}}$ is the vortex generation frequency.

In our recent experiments [4], this natural forcing universality was extended to periodically forced separated flow over stationary flat-plate wings. Pulsed dielectric-barrier-discharge (DBD) plasma perturbations at a frequency $f_p$ repeatedly severed the separated shear layer at the leading edge, generating discrete LEVs that convected downstream to form a train of vortices over the wing surface (Figure 2, left). For one case in which $h_b$ and $U_{\text{sep}}$ were measured directly, maximum lift occurred at $f_p D' / U_{\text{sep}} \approx 0.15$, consistent with the classical universal Strouhal range. Approximately two vortices were present over the wing surface at any instant in this maximum-lift range. However, direct measurement of $h_b$ and $U_{\text{sep}}$ was impractical for the remaining experimental conditions, motivating a generalized scaling applicable to the complete dataset.

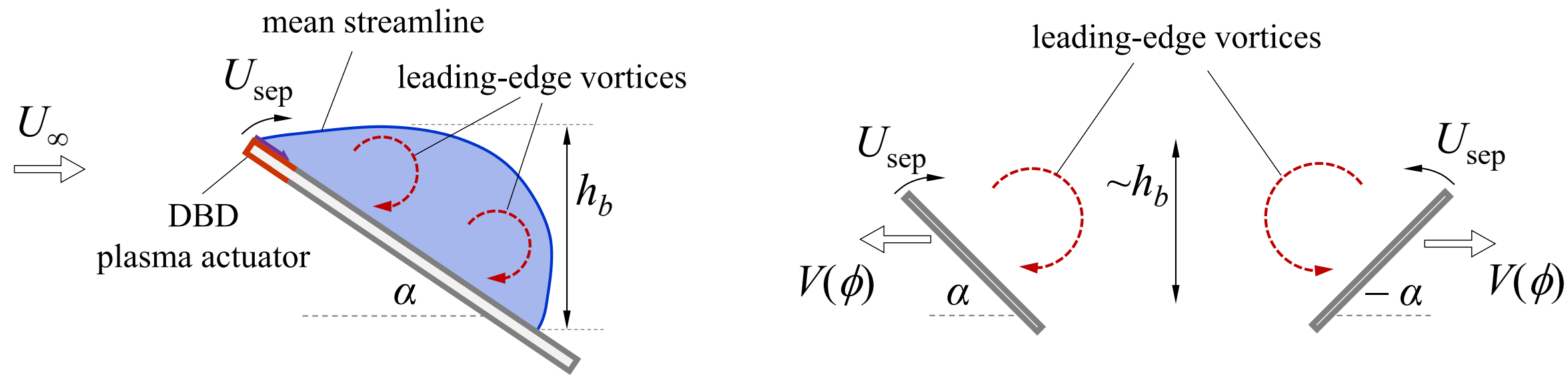

**Figure 2 | Correspondence between periodically forced and hovering-wing LEV generation.** (left) Periodic DBD perturbations of the leading-edge separated shear layer generate discrete LEVs that convect downstream, with approximately two vortices present over the stationary wing in the maximum-lift range [4,15]. (right) During hovering, one LEV of characteristic dimension $\sim h_b$ is generated during each translational half-stroke, giving two LEV-generation events per wingbeat. In both cases, the relevant aerodynamic frequency is the LEV-generation frequency.

The generalized model estimates the characteristic separated-flow dimension and separation velocity from wing geometry and aerodynamic loading. The separation-bubble height is related to the projected chord, $U_{\mathrm{sep}}$ is inferred from the upper-surface pressure using Roshko's relation [2], and lower-surface pressure is based on a modified Newtonian approximation [16]. A single calibration constant accounts for departures from these idealized pressure and geometric approximations; the complete development and sensitivity to these assumptions are given in the Supplementary Information.

Applying this model to the complete stationary-airfoil and wing datasets produces a substantially improved collapse of the maximum-lift conditions compared with conventional reduced-frequency scaling (Figure 3). Conditions corresponding to greater than 99% of the maximum normalized lift give $St_p = 0.171 \pm 0.018$, with approximately 10% relative scatter, compared with about 15% using reduced frequency. More importantly, the mean agrees closely with the classical scaling. The periodically forced stationary-wing experiments therefore indicate that maximum lift occurs when discrete LEVs are generated at approximately the classical vortex-shedding timescale.

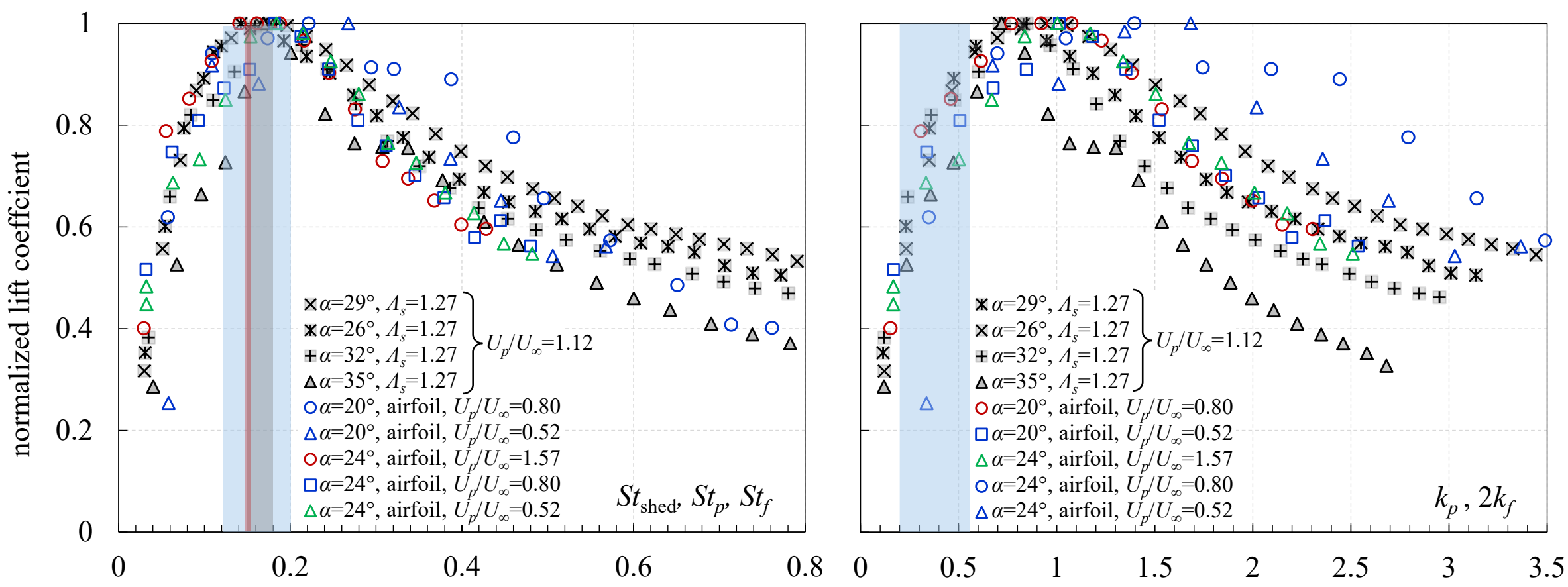

**Figure 3 | Strouhal scaling of periodically forced separated wings.** Normalized lift enhancement for periodically forced flat-plate wings and airfoils expressed in terms of (left): aerodynamic Strouhal number $St_p$; and (right) conventional reduced frequency $k_p$: $(1,200 \le Re \le 11,500)$. Blue shaded region: insect wing-flapping range based on $St_f$ (mean ± one standard deviation); grey shaded region Roshko's range for $St_{\mathrm{shed}}$ (mean ± one standard deviation) [2]; red line: average shedding value for wall-bounded flows from Sigurdson [3].

The key insight presented in this letter is that the stationary-wing experiments with artificially forced LEVs provide the physical link to hovering flight. Each DBD perturbation at frequency $f_p$ generates an LEV, making $f_p$ an LEV-generation frequency. During hovering, one LEV is similarly generated during each translational half-stroke (Figure 2, right). Two LEV-generation events therefore occur during each complete wingbeat, giving

$$f_{\text{LEV}} = 2f_f.$$

The correspondence between the two flows is therefore $f_p \leftrightarrow f_{\text{LEV}} = 2f_f$. This distinction is important: the common quantity is not the mechanical oscillation frequency, but the frequency at which discrete LEVs are generated.

We applied the same scaling to published hovering-flight data covering a wide range of insect morphologies and kinematics, including *Bombus terrestris* [17], *Manduca sexta* [18], *Coccinella septempunctata* [19,20], *Episyrphus balteatus* [19,21], *Eristalis tenax* [19,22], Apis mellifera [23], *Bombus hortorum* [19,21], *Bombus terrestris* [24], *Manduca sexta* [24], *Bombus ignitus* [25], *Macroglossum stellatarum* [26,27], and *Agrius convolvuli* [28]. The dataset spans $686 \le Re \le 7018$, nearly two orders of magnitude in body mass, and substantial variations in wing geometry, wingbeat frequency and stroke amplitude. The characteristic vortex dimension and separation velocity were determined using the same geometric and aerodynamic approximations developed for the stationary wings, using standard cycle-averaged quantities to provide representative values for the continuously varying wing kinematics. A representative aerodynamic angle of attack $\alpha'$ was determined separately for each dataset using the best available published hovering kinematics. Complete hovering-flight datasets that include the required morphology, kinematics and angle-of-attack, presented as Supplementary Information, are relatively scarce. Thus, the present database represents those published cases for which these quantities could be determined with reasonable confidence.

The aerodynamic Strouhal numbers for the hovering-insect datasets are shown as a function of Reynolds number in Figure 4 (left). The data exhibit relatively large scatter, as expected for biological data compiled from independent studies, reflecting natural interspecies variability as well as differences in experimental methods and reported kinematics. Despite this scatter, the Strouhal numbers cluster around $St_f = 0.162 \pm 0.038$ with no measurable dependence on Reynolds number. The mean lies close to both the stationary-wing value, $St_p = 0.171 \pm 0.018$, and Roshko's classical value of 0.167. Thus, independently generated LEVs—periodically forced over stationary wings and naturally generated during hovering— recover essentially the same characteristic Strouhal number over substantially different geometries, kinematics and Reynolds numbers. Note, however, that accounting for two LEV-generation events per wingbeat does not produce correspondence on the basis of reduced frequency, where $2k_f = 0.452 \pm 0.082$ (see Figure 3, right), compared to the stationary-wing maximum-lift value, $k_p = 0.885 \pm 0.135$. The agreement emerges only when the characteristic separated-flow dimension and separation velocity are introduced through the Strouhal scaling.

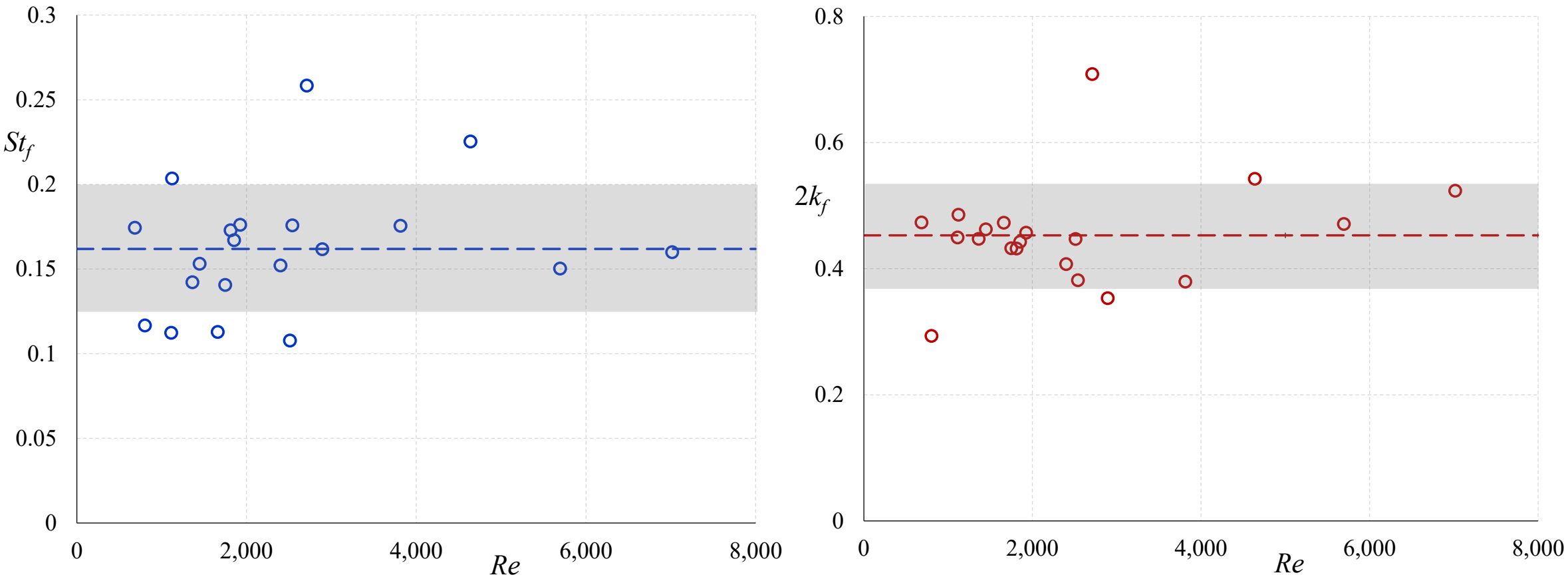

**Figure 4 | Aerodynamic frequency scaling of hovering insects.** (left): Aerodynamic Strouhal number $St_f$ and (right): twice the conventional reduced frequency $2k_f$, plotted against Reynolds number for the hovering-insect datasets [17-28]. Dashed lines indicate mean values and shaded regions denote ±1 standard deviation. The aerodynamic scaling gives $St_f = 0.162 \pm 0.038$, close to the stationary-wing and classical vortex-shedding values.

The principal uncertainty in the insect analysis is the representative angle of attack $\alpha'$. Wing incidence varies throughout the stroke and between up- and down-strokes, and is not consistently defined or reported in the literature. Furthermore, geometric incidence does not necessarily correspond to the effective aerodynamic angle relative to the local flow. Directly measured or reconstructed translational-stroke values were therefore used where available; otherwise, representative values were estimated from published species-specific kinematics. Despite this uncertainty, the relative scatter in $St_f$ is comparable to that of the directly determined $k_f$ shown in Figure 4 (right). Although both parameters are independent of wingbeat frequency, their physical interpretations are fundamentally different. Reduced frequency is determined solely by wing geometry and kinematics, whereas $St_f$ additionally incorporates the characteristic dimension and separation velocity of the LEV-forming flow. The close correspondence between $St_p = 0.171 \pm 0.018$, $St_f = 0.162 \pm 0.038$ and $St_{\text{shed}} = 0.167$ therefore points to a common vortex-formation timescale underlying periodically forced LEVs, naturally generated hovering LEVs and classical separated-flow vortex shedding.

The universal vortex-formation timescale also places a direct constraint on aerodynamic design. For a specified wing geometry and stroke amplitude, the scaling defines the aerodynamic loading compatible with sustained LEV formation. Maximum loading is predicted at representative angles of attack of approximately $57° - 64°$, whereas minimum aerodynamic power occurs at approximately $33° - 51°$. Furthermore, the mean value of $42°$ corresponds closely to the average representative angle of $40°$ obtained from hovering insects data considered here. Combined with the hovering force balance, the resulting loading determines the wingbeat frequency required to support the insect weight. Thus, the Strouhal scaling provides more than a frequency correlation: it connects the universal vortex-formation timescale to the wing kinematics, aerodynamic loading and power requirements of hovering flight.

These results indicate that the characteristic aerodynamic timescale of insect hovering is governed by LEV generation rather than simply by mechanical wing oscillation. Periodically forced stationary-wing LEVs and naturally generated hovering LEVs recover the classical 0.15 to 0.17 Strouhal number range despite substantial differences in geometry, kinematics and vortex-generation mechanisms. The emergence of the same Strouhal number in bluff-body wakes, separated-wing flows and insect hovering reveals a common vortex-formation timescale linking canonical separated flows to the aerodynamic mechanism that sustains biological flight.

– – – – – – – – – – – –

# Supplemental Information for:
# A Universal Vortex-Formation Law for Insect Hovering

David Greenblatt

Technion – Israel Institute of Technology, Technion City, Haifa 3200003, Israel

## Nomenclature

### Roman Symbols

| | |
|---|---|
| $A$ | peak-to-peak wingtip flapping amplitude (m) |
| $c$ | airfoil or wing chord length (m) |
| $c_m$ | insect mean wing chord length (m) |
| $C_L$ | stationary wing lift coefficient, $F_L / \frac{1}{2}\rho U_\infty^2 Rc$ |
| $\bar{C}_L$ | cycle-averaged lift coefficient, $\bar{F}_L / \frac{1}{2}\rho \bar{V}^2 Rc_m$ |
| $C_D$ | stationary wing drag coefficient, $F_D / \frac{1}{2}\rho U_\infty^2 Rc$ |
| $\bar{C}_D$ | cycle-averaged drag coefficient, $\bar{F}_D / \frac{1}{2}\rho \bar{V}^2 Rc_m$ |
| $C_1$ | empirical calibration constant |
| $\hat{C}_{p,l}$ | chord-averaged lower-surface pressure coefficient |
| $\hat{C}_{p,u}$ | chord-averaged upper-surface pressure coefficient |
| $D$ | body dimension normal to $U_\infty$ (m) |
| $D'$ | characteristic bluff-body wake dimension (m) |
| $f_f$ | wingbeat (or flapping) frequency (Hz) |
| $f_p$ | perturbation frequency (Hz) |
| $f_{\text{shed}}$ | vortex-shedding frequency (Hz) |
| $F$, $\bar{F}$ | separation-velocity scaling functions |
| $\bar{F}_L$ | cycle-averaged lift force (N) |
| $\bar{F}_D$ | cycle-averaged drag force (N) |
| $g$ | gravitational acceleration (m/s$^2$) |
| $h_b$ | mean separation-bubble height, $D'/2$ (m) |
| $k_f$ | flapping reduced frequency, $\pi f_f c_m / \bar{V}$ |
| $k_p$ | perturbation reduced frequency, $\pi f_p c / U_\infty$ |
| $m$ | insect or vehicle mass (kg) |
| $P_{\text{aero}}$ | aerodynamic power (W) |
| $R$ | wing semispan (m) |
| $Re_f$ | flapping-wing Reynolds number, $\bar{V} c_m / \nu$ |
| $Re$ | bluff-body Reynolds number, $U_\infty D / \nu$ |
| $St_b$ | biomechanics Strouhal number, $f_f A / V_{\text{ref}}$ |
| $St_f$ | wing-flapping aerodynamic Strouhal number, $(2f_f)D' / \bar{V}_{\text{sep}}$ |
| $St_p$ | perturbation aerodynamic Strouhal number, $f_p D' / U_{\text{sep}}$ |
| $St_{\text{shed}}$ | universal bluff-body vortex-shedding Strouhal number, $f_{\text{shed}} D' / U_{\text{sep}}$ |
| $U_\infty$ | freestream speed (m/s) |
| $U_p$ | DBD plasma perturbation amplitude (m/s) |
| $U_{\text{sep}}$ | wind-speed at separation (m/s) |
| $V(\phi)$ | phase-dependent wing-tip speed (m/s) |
| $\bar{V}$ | cycle-averaged wing-tip speed (m/s) |
| $\bar{V}_{\text{sep}}$ | cycle-averaged flapping-wing speed at separation (m/s) |
| $V_{\text{ref}}$ | reference speed used in $St_b$ (m/s) |

**Greek Symbols**

| | |
|---|---|
| $\alpha$ | angle-of-attack (°) |
| $\alpha'$ | representative angle-of-attack (°) |
| $\phi$ | phase-angle within the flapping cycle (°) |
| $\Phi$ | wingtip peak-to-peak flapping amplitude (°) |
| $\Lambda_s$ | semispan aspect ratio, $R / c$ or $R / c_m$ |
| $\nu$ | air kinematic viscosity ($m^2$/s) |
| $\rho$ | air density (kg/$m^3$) |

# 1 Introduction

Wingbeat frequency is a critical aspect of flapping-wing micro aerial vehicle design and is based largely on insect-flight scaling laws. For hovering in the $\mathcal{O}(10^2)$ to $\mathcal{O}(10^4)$ Reynolds-number range, lift is generated primarily through leading-edge vortices (LEVs), arising from the alternating motion of the wings (Figure 1) [1,2]. The central role of LEVs in insect flight was established in the classical experiments of Ellington and co-workers [3], and subsequently through dynamically scaled experiments identifying delayed stall, rotational circulation and wake capture as important unsteady aerodynamic mechanisms [4]. The persistence and attachment of the LEV during the translational portion of the stroke have since been investigated extensively [5], and LEV-mediated lift is now recognized as a defining feature of hovering insect flight.

Flapping-wing kinematics are commonly characterized using either the biomechanics Strouhal number $St_b \triangleq f_f A / V_{\text{ref}}$ [6] or the reduced frequency $k_f \triangleq \pi f_f c_m / V_{\text{ref}}$ [7-9]. The biomechanics Strouhal number has proven to be useful for correlating cruising animal flight [6] but is of limited use for hovering because $V_{\text{ref}}$ is itself proportional to wingbeat frequency and the characteristic wing excursion. Similarly, substitution of $\bar{V} = 2 f_f \Phi R$ into the reduced frequency definition gives $k_f = \pi / 2\Phi\Lambda_s$, where $\Lambda_s = R / c_m$ is the semispan aspect ratio. Hence $f_f$ again cancels, and $k_f$ corresponds to the range approximately 0.2–0.4 for a wide variety of insects. Moreover, $k_f$ is borrowed from classical aerodynamics studies, and simply represents the ratio of convective to oscillatory timescales [10,11]. It therefore has no direct physical linkage to the periodic formation and shedding of LEVs.

The aerodynamic Strouhal number considered here is fundamentally different from the biomechanics Strouhal number. Rather than being based on flapping amplitude and a reference flight velocity, it is based on the characteristic vortex dimension and separation velocity and therefore directly characterizes vortex formation and convection. Although insect-wing LEVs are associated with vortex shedding [2], a direct scaling relationship between lift forces and vortex-shedding frequencies has not been explicitly identified. In the arguments below, it is shown that insect hovering and bluff-body vortex shedding are governed by the same universal aerodynamic Strouhal-number scaling.

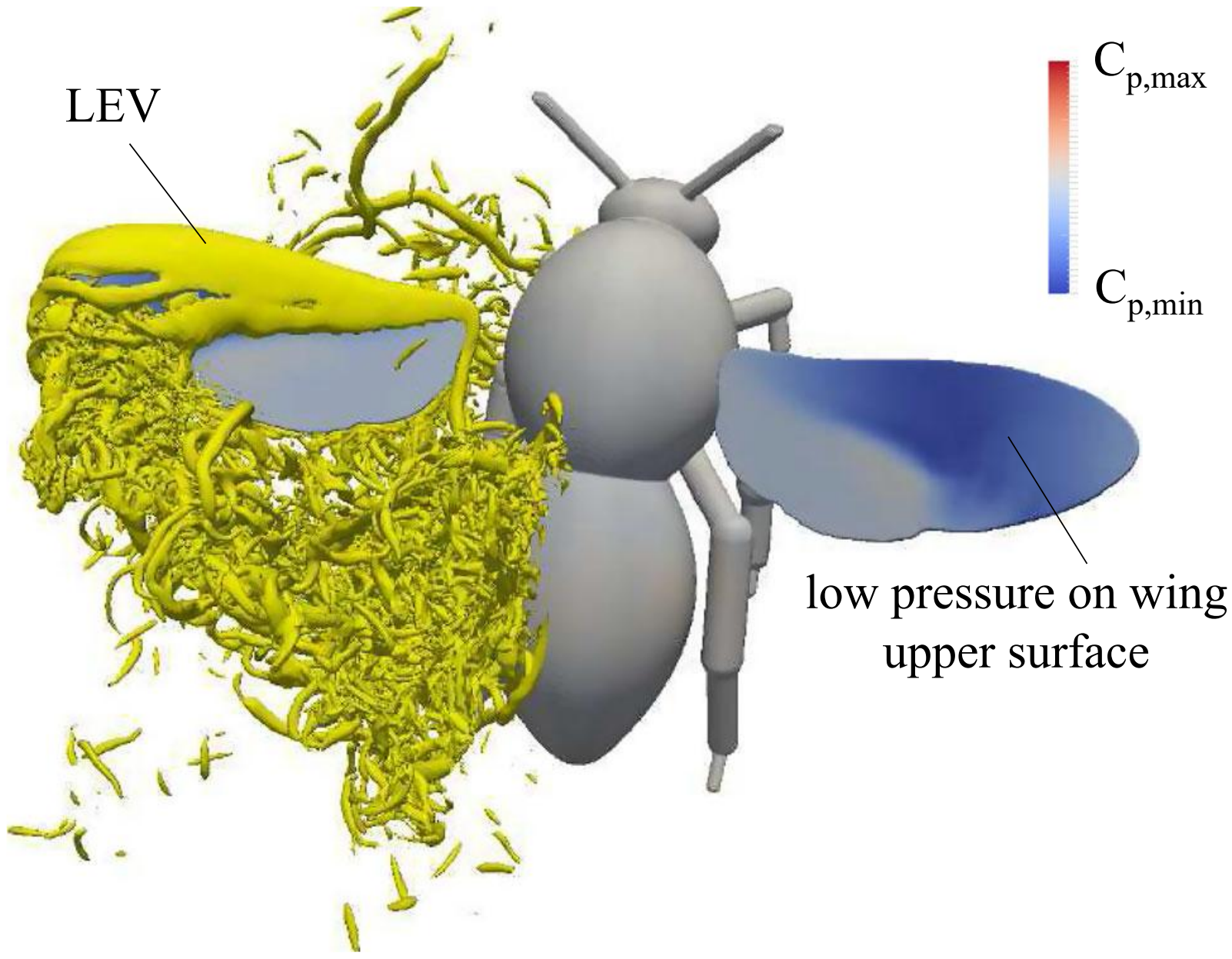

Figure 1. Results from a numerical simulation of a hovering bumblebee, with flapping and revolving wings, illustrating the effect of the LEV by means of the Q-criterion isosurface (left wing) and the pressure coefficient (right wing). From [12], republished with permission.

# 2 Forced Vortex Shedding

## 2.1 The 'Universal' Strouhal number

Two-dimensional bluff-body vortex shedding, irrespective of geometry, is characterized by Roshko's 'universal' Strouhal number [13]:

$$St_{\text{shed}} \triangleq \frac{f_{\text{shed}} D'}{U_{\text{sep}}} = 0.167 \text{ (mean)} \pm 0.011 \text{ (standard deviation)} \tag{1}$$

which is valid when the separating shear layer undergoes transition to turbulence, i.e., for $400 \lesssim Re_D \triangleq U_\infty D / \nu \lesssim 20{,}000$ (see Appendix A) [13]. A major advantage of this scaling is that the drag coefficient on an arbitrary bluff body can be calculated by a combination of potential flow theory and the assumption that the base pressure and velocity at separation are related by $C_{pb} = 1 - (U_{\text{sep}} / U_\infty)^2$. Equally significant, Sigurdson [14] compiled natural shedding frequencies from a wide range of separated and reattaching flows and showed that the global separation-bubble mode scales over the relatively narrow range $0.07 \leq fh_b / U_{\text{sep}} \leq 0.08$. Taking the characteristic wake dimension as $D' = 2h_b$ gives $St_{\text{shed}} \approx 0.14$ to 0.16, i.e., centered around 0.15. Sigurdson thereby extended the universal Strouhal scaling to wall-bounded separated flows. Together with Roshko's value of $St_{\text{shed}} \approx 0.167$, these results define an approximately universal vortex-shedding range of $0.15 \lesssim St_{\text{shed}} \lesssim 0.17$. Here, $D'$ represents the cross-wake dimension, while $h_b = D'/2$ provides the characteristic vortex dimension.

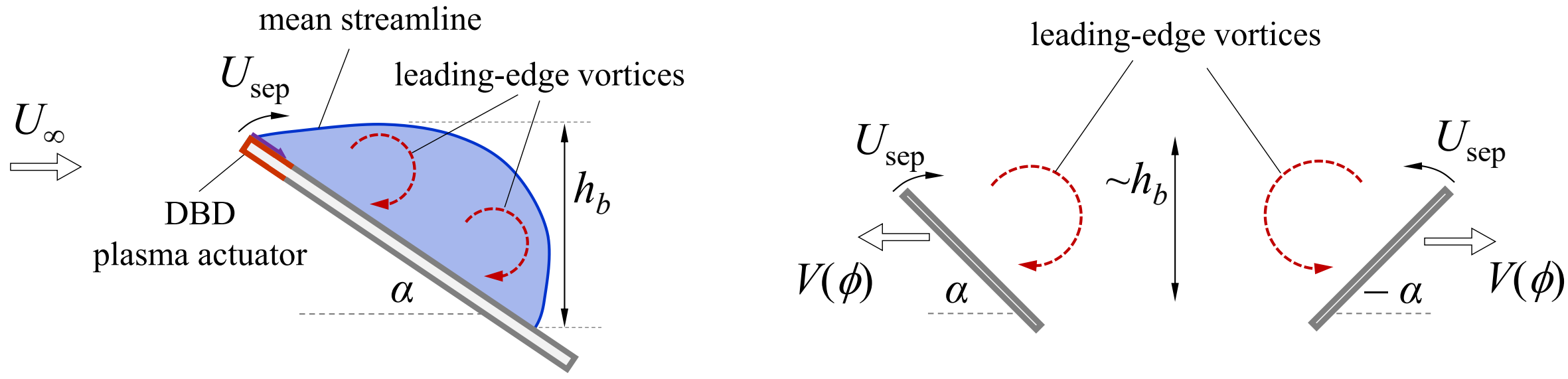

Figure 2. Correspondence between periodically forced LEV generation on a stationary separated wing and LEV generation during hovering wing motion. (left) Periodic DBD perturbations of the leading-edge separated shear layer generate discrete LEVs that convect downstream, with approximately two vortices present over the stationary wing in the maximum-lift range [15,16]. (right) During hovering, one LEV of characteristic dimension $h_b$ is generated during each translational half-stroke, giving two LEV-generation events per wingbeat. In both cases, the relevant aerodynamic frequency is the LEV-generation frequency.

In our recent study [15], we extended this framework in two respects. First, discrete LEVs were generated deliberately using pulsed DBD plasma forcing, with the plasma perturbation frequency $f_p$ replacing the natural vortex-shedding frequency $f$ (see the schematic in Figure 2, left). Second, the maximum-lift condition occurred at $f_p D' / U_{\text{sep}} \approx 0.15 \pm 0.01$, in close agreement with the universal Strouhal scaling. The perturbations repeatedly sever the leading-edge separated shear layer [16], generating discrete LEVs that convect downstream to produce a train of vortices over the wing surface, with the presence of approximately two vortices on the surface at any instant corresponding to the maximum-lift range (also shown schematically in Figure 2, left). This latter point is significant in the context of hovering flight, where two LEVs are generated during each wingbeat, one during each translational half-stroke (shown schematically in Figure 2, right). In the analysis below, we first extend the forced-vortex scaling to the complete stationary-airfoil and wing datasets for which $h_b$ and $U_{\text{sep}}$ cannot be measured directly. The resulting generalized scaling is then applied to hovering insect flight to determine whether naturally generated flapping-wing LEVs follow the same maximum-lift scaling as the periodically forced LEVs. The analysis is limited to the Reynolds number range cited above, for which the universal Strouhal number is valid.

## 2.2 Universal Flat-Plate Scaling

Accurate determination of $U_{\text{sep}}$ and $h_b$ requires detailed flow-field measurements [15], which is not practical for the multiple data sets. Instead, we develop an approach that requires a single calibration constant, based on the chord-averaged upper and lower surface pressure coefficients: $C_L = (\hat{C}_{p,u} - \hat{C}_{p,l})\cos\alpha$. For the separated flat-plate wings, the chord-averaged lower-surface pressure is represented by the modified Newtonian form $\hat{C}_{p,l} = C_1 \sin^2\alpha$ [17], while the upper-surface pressure is related to the separation velocity through Roshko's relation $\hat{C}_{p,u} \approx 1 - (U_{\text{sep}} / U_\infty)^2$. The separation-bubble height is similarly approximated as $h_b \approx c \sin\alpha$. Thus, $C_1 = O(1)$ accounts collectively for departures from these idealized pressure and geometric approximations. Guided by the directly measured stationary-wing result, $f_p D' / U_{\text{sep}} \approx 0.15 \pm 0.01$, the simple choice $C_1 = 1$ was adopted and subsequently held fixed for all stationary-wing and insect analyses.

Figure 3 shows normalized lift coefficient results plotted on the basis of reduced frequency $k_p$ and

$$St_p \triangleq \frac{f_p D'}{U_{\text{sep}}} \approx \frac{f_p c}{U_\infty}\frac{2\sin\alpha}{F} = \frac{2k_p \sin\alpha}{\pi F} \tag{2}$$

where $F = (\cos^2\alpha + C_L / \cos\alpha)^{1/2}$. The Strouhal number scaling shows a far more convincing collapse of the data, particularly in the angle of attack range 24° to 35°, where the data points greater than 99% of the normalized lift coefficient yield $k_p = 0.885 \pm 0.135$ (15%) and $St_p = 0.171 \pm 0.0180$ (10%). Crucially, the averages of $St_p$ and Roshko's $St_{\text{shed}}$ compare favorably as shown by the grey shaded region in Figure 3 (left) with a corresponding mean value of 0.167 (see Appendix A). Physically, each leading-edge perturbation velocity $U_p$ severs the shear layer, leading to a train of two vortices over the wing surface at any instant and, in a mean sense, a large vortex of height $h_b \approx c\sin\alpha$ is 'trapped' above the surface. The superior data collapse for $St_p$ versus $k_p$ scaling, together with LEV-mediated lift mechanism for this simplified problem, suggests that similar scaling may provide a physically-based vortex-formation similarity parameter for insect hovering.

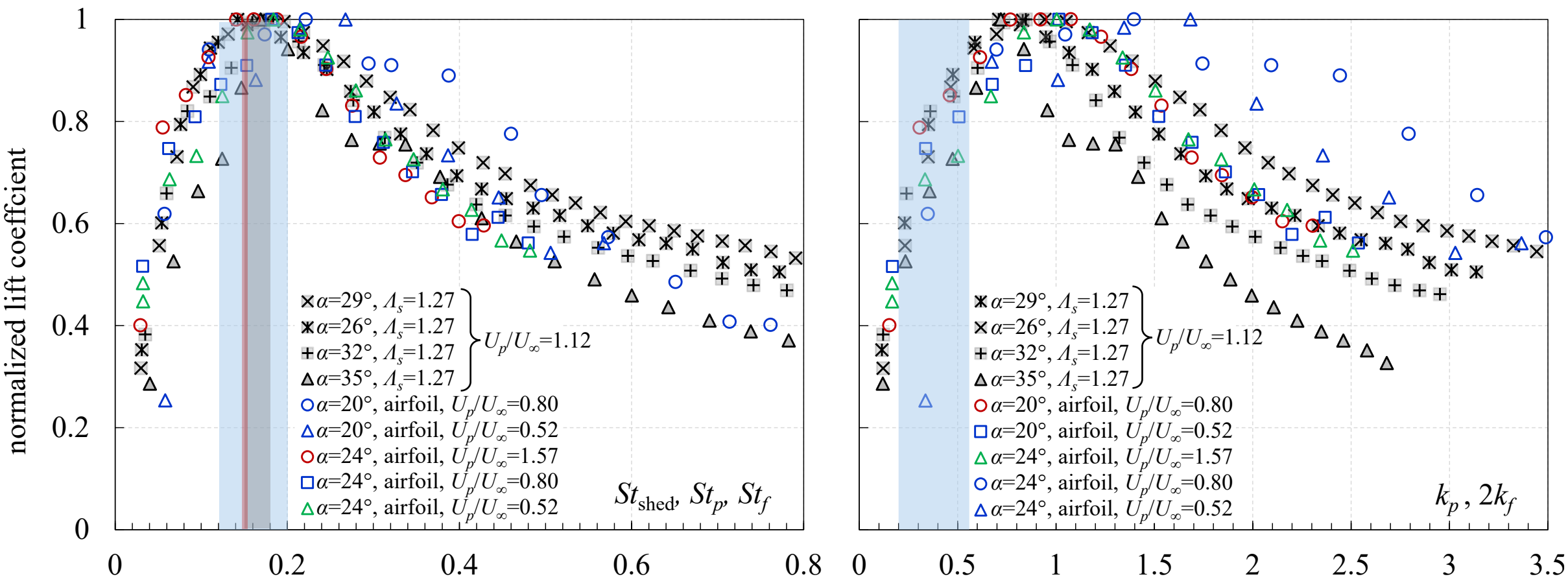

Figure 3. Normalized changes in post-stall lift coefficient for flat-plate semispan wings and flat-plate airfoils based on universal Strouhal-number (left) and reduced-frequency (right) scaling $(1,200 \le Re \le 11,500)$. Blue shaded region: insect wing-flapping range (mean ± one standard deviation); grey shaded region: Roshko's bluff-body range (mean ± one standard deviation) [13]; red line: mean value for wall-bounded separated flows from Sigurdson [14].

The choice of placing the lumped correction through $C_1$ in the lower-surface pressure relation is not unique, and an analogous correction could instead be introduced into the upper-surface pressure coefficient relation. For example, at representative values $\alpha = 30°$ and $C_L = 1$, expansion about unity gives $\Delta St_p / St_p \approx 0.066 \Delta C_1$ for the present formulation, whereas an equivalent multiplicative correction to the upper-surface velocity relation, namely $\hat{C}_{p,u} \approx 1 - C_2 (U_{\text{sep}} / U_\infty)^2$, would give $\Delta St_p / St_p \approx 0.5 \Delta C_2$. Thus, the numerical sensitivity depends on where the lumped order-unity correction is introduced, and $C_1 = 1$ should be regarded as a simple closure rather than a uniquely determined calibration.

## 3 Hovering Insects

The close correspondence between the stationary-wing results and Roshko's classical bluff-body scaling provides the basis for extending the same framework to hovering insect wings. Importantly, the stationary-wing Strouhal number is based on the perturbation frequency $f_p$, with each perturbation generating an LEV. Thus, the relevant frequency is fundamentally a vortex-generation frequency rather than a mechanical oscillation frequency. An important difference is that the corresponding quantities must be defined on a cycle-averaged basis to provide representative values for the continuously varying wing kinematics. In particular, we use the cycle-averaged lift coefficient $\bar{C}_L = mg / (\frac{1}{2} \rho \bar{V}^2 R c_m)$ and wing-tip speed $\bar{V}$ [2]. A representative angle of attack $\alpha'$ must also be identified because $\alpha$ varies substantially throughout the stroke. These quantities are then used with the same pressure and geometric approximations developed for the stationary wings, retaining $C_1 = 1$ without further adjustment.

For the present scaling, the relevant frequency is taken to be the LEV-generation frequency. This definition follows directly from the stationary-wing experiments, where each leading-edge perturbation at $f_p$ generates an LEV. In hovering flight, one LEV is similarly generated during each translational half-stroke, so that two LEV-generation events occur per wingbeat and the corresponding aerodynamic frequency is $f_{\text{LEV}} = 2 f_f$. Thus, the direct correspondence between the stationary and flapping-wing problems is $f_p = f_{\text{LEV}} = 2 f_f$, so that the wing-flapping Strouhal number is defined as:

$$St_f \triangleq \frac{(2 f_f) D'}{\bar{V}_{\text{sep}}} \approx \frac{2 k_f}{\pi} \frac{2 \sin \alpha'}{\bar{F}} = \frac{2 \sin \alpha'}{\Lambda_s \Phi \bar{F}} \tag{3}$$

where $\bar{F} = (\cos^2 \alpha' + \bar{C}_L / \cos\alpha')^{1/2}$ . A summary of hovering-insect data, with the result of equation (3), are shown in Table 1. The data encompass a broad range of insect morphologies and hovering kinematics, spanning more than an order of magnitude in Reynolds number $(686 \le Re_f \le 7{,}018)$, approximately two orders of magnitude in mass, and large variations in wing geometry, wingbeat frequency and stroke amplitude. Despite these differences, the Strouhal numbers cluster around $St_f = 0.162 \pm 0.038$ (Figure 4, left), and exhibit only a very weak dependence on Reynolds number ($\mathrm{R}^2 = 0.055$). Most significantly, this mean lies within the Roshko range identified by the grey shaded region in Figure 3 (left).

Table 1: Morphological, kinematic and aerodynamic parameters of the hovering-insect datasets used to evaluate the proposed Strouhal scaling; representative angles of attack $\alpha'$ and their determination are detailed in Table B1.

| Ref. | species/study | $m$ (mg) | $\phi$ (deg) | $f$ (Hz) | $R$ (mm) | $c_m$ (mm) | $\Lambda_s$ | $\bar{V}$ (m/s) | $Re_f$ | $\alpha'$ | $\bar{C}_L$ | $k_f$ | $St_f$ |
|---|---|---|---|---|---|---|---|---|---|---|---|---|---|
| [18] | Bombus terrestris | 175.0 | 120 | 130 | 13.2 | 4.0 | 3.28 | 7.2 | 1926 | 42.5 | 0.52 | 0.228 | 0.176 |
| | | 180.0 | 120 | 120 | 13.7 | 4.0 | 3.39 | 6.9 | 1855 | 42.5 | 0.56 | 0.221 | 0.167 |
| [19] | Manduca sexta | 1500.0 | 120 | 25 | 51.0 | 16.0 | 3.19 | 5.3 | 5697 | 35.0 | 0.52 | 0.235 | 0.150 |
| [20] | Coccinella septempunctata | 34.4 | 177 | 54 | 11.2 | 3.2 | 3.47 | 3.7 | 805 | 45.0 | 0.55 | 0.147 | 0.117 |
| [20] | Episyrphus balteatus | 27.3 | 90 | 160 | 9.3 | 2.2 | 4.23 | 4.7 | 686 | 40.0 | 0.50 | 0.237 | 0.174 |
| [20] | Eristalis tenax | 68.4 | 109 | 157 | 11.4 | 3.2 | 3.57 | 6.8 | 1448 | 33.0 | 0.33 | 0.231 | 0.153 |
| [20] | Apis mellifera | 101.9 | 131 | 197 | 9.8 | 3.1 | 3.18 | 8.83 | 1813 | 40.0 | 0.351 | 0.216 | 0.173 |
| [21] | Eristalis tenax | 88.9 | 107 | 164 | 11.2 | 3.0 | 3.76 | 6.9 | 1364 | 33.8 | 0.46 | 0.224 | 0.142 |
| | | 165.9 | 110 | 209 | 11.2 | 2.9 | 3.80 | 8.9 | 1751 | 35.5 | 0.51 | 0.216 | 0.141 |
| | | 142.7 | 109 | 142 | 10.7 | 2.9 | 3.67 | 5.8 | 1113 | 34.2 | 1.13 | 0.225 | 0.112 |
| [20] | Bombus hortorum | 226.0 | 120 | 152 | 14.1 | 4.2 | 3.36 | 9.0 | 2514 | 25.0 | 0.38 | 0.223 | 0.108 |
| [22] | Apis mellifera | 100.0 | 91 | 230 | 9.7 | 2.4 | 4.08 | 7.1 | 1124 | 55.0 | 0.70 | 0.243 | 0.203 |
| [23] | Bombus terrestris | 175.0 | 116 | 116 | 13.2 | 4.0 | 3.28 | 6.2 | 1662 | 28.0 | 0.70 | 0.236 | 0.113 |
| [23] | Manduca sexta | 1648.0 | 121 | 26 | 51.9 | 18.3 | 2.84 | 5.8 | 7018 | 32.0 | 0.42 | 0.262 | 0.160 |
| [24] | Bombus ignitus | 430.0 | 139 | 145 | 15.0 | 4.1 | 3.66 | 10.6 | 2895 | 52.0 | 0.51 | 0.176 | 0.162 |
| | | 413.0 | 121 | 132 | 15.3 | 4.2 | 3.64 | 8.6 | 2402 | 46.0 | 0.71 | 0.204 | 0.152 |
| | | 588.0 | 130 | 138 | 18.2 | 5.0 | 3.64 | 11.5 | 3816 | 48.0 | 0.40 | 0.190 | 0.176 |
| | | 248.0 | 129 | 145 | 14.6 | 4.0 | 3.65 | 9.5 | 2541 | 47.0 | 0.38 | 0.191 | 0.176 |
| [25] | Macroglossum stellatarum | 281.0 | 95 | 74 | 21.0 | 7.9 | 2.68 | 5.2 | 2711 | 40.0 | 0.52 | 0.354 | 0.258 |
| [26] | Agrius convolvuli | 956.0 | 115 | 39 | 35.8 | 12.4 | 2.89 | 5.6 | 4641 | 47.5 | 0.55 | 0.271 | 0.225 |

For comparison, the conventional reduced frequency is plotted against Reynolds number in Figure 4 (right) and exhibits essentially the same statistical behavior, with $k_f = 0.226 \pm 0.041$ and a similarly negligible Reynolds number dependence ($\mathrm{R}^2 = 0.056$). This is not unexpected, since $k_f$ is determined directly from the flapping kinematics. The relatively large scatter in both data sets is expected, because these data are drawn from different insect species and experimental methodologies, rather than from carefully controlled wind-tunnel experiments. The determination of $St_f$, however, additionally requires estimates of the characteristic vortex dimension and separation velocity, both of which depend on the representative angle of attack $\alpha'$. Despite the substantial uncertainty associated with estimating $\alpha'$ from published insect kinematics (see below and Appendix B), the percentage scatter in $St_f$ (Figure 4, left; $\sigma_{St_f} / \overline{St}_f = 23\%$) is comparable to that in $k_f$ (Figure 4, right; $\sigma_{k_f} / \overline{k}_f = 18\%$).

It is important to note that accounting for two LEV-generation events per wingbeat does not produce correspondence on the basis of reduced frequency. This is shown by comparing the blue shaded regions corresponding to $St_f = 0.162 \pm 0.038$ and $2k_f = 0.452 \pm 0.082$ in Figure 3 (left and right, respectively), where the latter is well below the stationary-wing maximum-lift value, $k_p = 0.885 \pm 0.135$. The agreement emerges only when the characteristic separated-flow dimension and separation velocity are introduced through the Strouhal scaling.

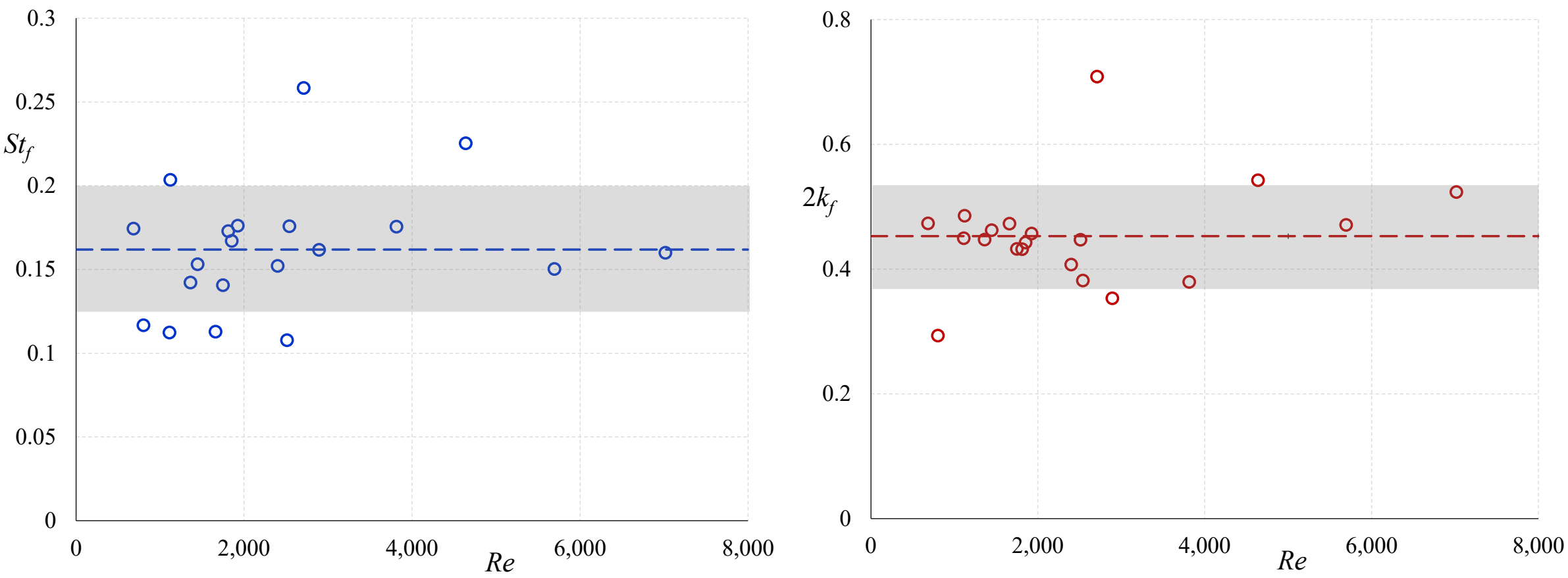

Figure 4. Aerodynamic frequency scaling for hovering insects. Aerodynamic Strouhal number $St_f$ (left) and twice the conventional reduced frequency $2k_f$ (right) as functions of Reynolds number for the hovering-insect datasets in Table 1. Dashed lines indicate the respective mean values, and grey bands denote the corresponding ± 1 standard-deviation ranges.

A major source of uncertainty in applying Strouhal number scaling to insects is the determination of $\alpha'$. Unlike the stationary-wing experiments, insect wing angles of attack vary within the stroke and between up- and down-strokes. Furthermore, they are not consistently reported in the literature. In addition, published values often represent geometric wing incidence rather than the effective aerodynamic angle of attack relative to the local flow. Consequently, $\alpha'$ was determined individually for each data set from the best available hovering-flight kinematics, as summarized in Appendix B. Where possible, directly reported or reconstructed translational-stroke values were used; otherwise, representative values were estimated from published species-specific wing kinematics. Thus, $\alpha'$ should not be interpreted as a uniformly defined cycle-averaged quantity. Nevertheless, despite this being the principal source of uncertainty in the analysis, the resulting scatter in $St_f$ remains comparable to that of the directly determined reduced frequency $k_f$.

The persistence of LEVs on flapping insect wings has been attributed to three-dimensional stabilization mechanisms, including spanwise flow and vorticity transport driven by spanwise pressure gradients [5,30-32]. Such mechanisms can limit LEV growth and prolong attachment during the translational portion of the stroke. They do not, however, preclude a common timescale governing LEV formation. The present Strouhal scaling concerns the generation and characteristic dimension of the LEV rather than its subsequent residence time or attachment. Thus, three-dimensional spanwise transport may govern LEV stabilization after formation, while the close correspondence between $St_f$ , $St_p$ and classical vortex-shedding values suggests that the characteristic formation timescale remains governed by the same underlying scaling.

## 4 Aerodynamic Design Considerations

In addition to providing the appropriate physical scaling for LEV formation and convection, the Strouhal number has practical advantages over the conventional reduced frequency. Specifically, $k_f$ does not provide a relationship between kinematics and aerodynamic loading. To show this, the Strouhal scaling expressed in equation (3), is rearranged to give:

$$\bar{C}_L = \cos\alpha'\left[(2\sin\alpha' / St_f \Lambda_s \Phi)^2 - \cos^2\alpha'\right]. \quad (4)$$

Thus, for specified $\Lambda_s$ and $\Phi$, adoption of the universal vortex-shedding value $St_f \approx 0.16$ uniquely defines a $\bar{C}_L(\alpha)$ relationship, with the peak at:

$$\alpha'_{\max} = \cos^{-1}\left[3\left(1+(St_f \Lambda_s \Phi / 2)^2\right)\right]^{-1/2} \quad (5)$$

For the insect geometries and stroke amplitudes shown in Table 1, this relation gives $57° \le \alpha'_{\max} \le 64°$, with a mean of about $60°$. This range is higher than the range shown in Table 1, although it does not imply that hovering insects should operate at this condition. Rather, the universal-$St_f$ relation defines the maximum aerodynamic loading compatible with the preferred vortex-formation timescale, whereas the energetic cost of producing this loading increases strongly with $\alpha$. To illustrate this, the observed relationship $C_D = C_L \tan\alpha$ on stationary wings [15] is assumed to apply to cycle-averaging on insect wings as $\bar{C}_D = \bar{C}_L \tan\alpha'$, which means that the drag penalty at higher angles. A first-order aerodynamic-power estimate, $P_{\text{aero}} \sim \bar{F}_D \bar{V}$, combined with the hovering force balance and equation (4), gives $P_{\text{aero}} \propto \tan\alpha' / \sqrt{\bar{C}_L}$. Minimization of this relation (see Appendix C) gives $33° \le \alpha'_{\text{opt}} \le 51°$ for the insect geometries shown in Table 1, with a mean value of $42°$ that corresponds well with the average value of $40°$. It is important to note that these angles are not fitted to the insect kinematics; rather, they emerge from minimizing the aerodynamic power requirement subject to the universal Strouhal-number condition. The fact that the minimum-power condition occurs well below the maximum-loading condition, suggests that hovering insects operate at lower angles as a compromise between aerodynamic loading and the power required to generate it.

This formulation has a potentially useful practical consequence for the design of hovering flapping-wing robots. For prescribed wing geometry and stroke amplitude, the universal-$St$ condition provides a target $\bar{C}_L$ as a function of $\alpha'$. Then, rearranging the hovering force balance, $mg = 2\rho f^2 \Phi^2 R^3 c_m \bar{C}_L$, gives an expression for the wingbeat frequency:

$$f_f = \left( \frac{mg}{2\rho \Phi^2 R^3 c_m \bar{C}_L} \right)^{1/2} . \qquad (6)$$

Thus, for prescribed vehicle mass, wing geometry and stroke amplitude, the universal-$St_f$ condition defines the aerodynamic loading and efficiency as a function of $\alpha'$ [equations (4), (5) and (9)], while the hovering force balance determines the corresponding wingbeat frequency [equation (6)]. Maximum loading minimizes the required wingbeat frequency, while the minimum-power condition selects a lower angle of attack and correspondingly higher frequency. This framework provides a relationship between vortex dynamics, aerodynamic loading, energetic cost and the wingbeat frequency required for hovering.

# 5 Conclusions

The results presented here extend the vortex-shedding scaling established for periodically forced separated flows to naturally generated LEVs in hovering insect flight. For stationary separated wings, periodic leading-edge perturbations generate discrete LEVs, and extension of the scaling to the complete airfoil and wing datasets gives a maximum-lift condition of $St_p = 0.171 \pm 0.018$, closely matching Roshko's classical value of approximately 0.167. In these experiments, the perturbation frequency $f_p$ is directly the LEV-generation frequency. In hovering flight, one LEV is generated during each translational half-stroke, giving $f_{\text{LEV}} = 2f$. Applying the same scaling to published hovering-insect data, without further adjustment of the calibrated pressure/separation model, gives $St_f = 0.162 \pm 0.038$ across a broad range of morphologies, kinematics and Reynolds numbers. The close agreement between the stationary-wing and insect results suggests that periodically forced and naturally generated LEVs are governed by the same characteristic vortex-formation timescale, consistent with the classical universal Strouhal-number range.

The principal uncertainty in applying the scaling to hovering insects is the representative angle of attack $\alpha'$, which is not consistently defined or reported and varies substantially throughout the wing stroke. Nevertheless, the percentage scatter in $St_f$ remains comparable to that of the directly determined reduced frequency $k_f$, despite the additional uncertainty associated with estimating the characteristic vortex dimension and separation velocity. More importantly, unlike $k_f$, $St_f$ incorporates these physically relevant vortex scales and independently recovers the classical universal Strouhal-number range.

The universal-$St_f$ condition also has direct aerodynamic and design implications. For prescribed wing geometry and stroke amplitude, it defines the aerodynamic loading as a function of $\alpha'$. Maximum loading occurs at relatively high angles and minimizes the wingbeat frequency required for hover, whereas inclusion of the aerodynamic drag penalty predicts a lower-angle minimum-power condition. Together with the hovering force balance, the framework therefore provides a direct connection between LEV-generation dynamics, aerodynamic loading, energetic cost and the wingbeat frequency required for hovering, and offers a physically based criterion for the analysis and design of hovering flapping-wing systems.

## Appendix A: Classical bluff-body Strouhal numbers

The Strouhal numbers compiled by Roshko [13] for vortex shedding from a broad range of two-dimensional bluff bodies, are presented in Figure A1. Despite substantial differences in body geometry, the data exhibit a pronounced concentration around 0.155 to 0.17, with an overall mean of $\overline{St} = 0.167$. The spread reflects the influence of geometry and associated differences in the separated shear layers, while the clustering demonstrates the relative insensitivity of the characteristic shedding frequency when scaled using the appropriate separation velocity and wake dimension. This classical result provides the reference against which the separated-wing and hovering-insect results are compared in the present study.

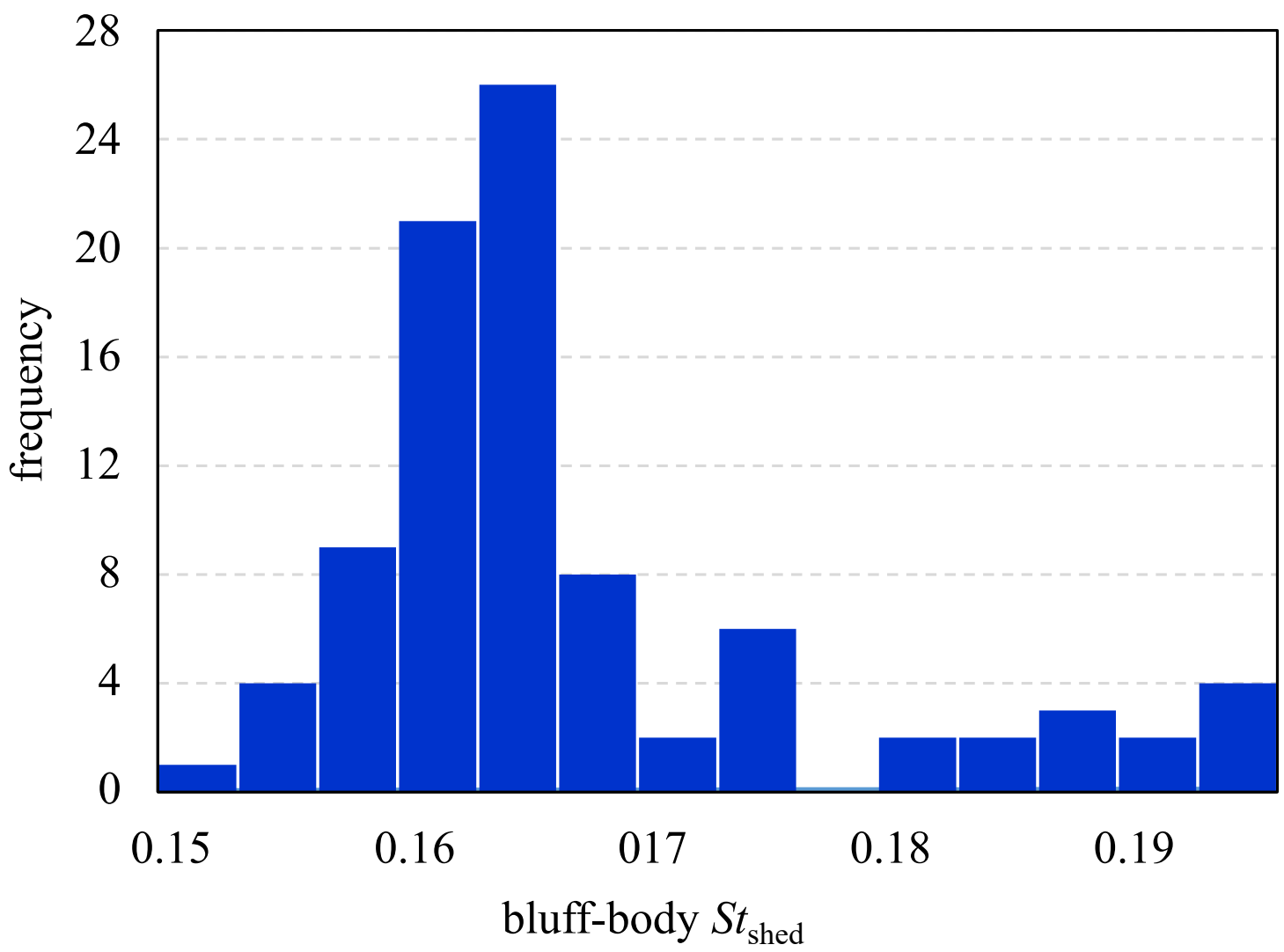

Figure A1. Classical bluff-body Strouhal numbers frequency distribution [13] obtained for two-dimensional bluff bodies comprising circular cylinders, normal flat plates and 90° wedges, over the Reynolds-number range $400 \lesssim Re \lesssim 20,000$. The mean and median values are 0.167 and 0.164 respectively.

## Appendix B: Angle of Attack Determination and Provenance

The representative angle of attack $\alpha'$ constitutes the principal uncertainty in determining $St_f$, since insect wing incidence varies throughout the stroke and published studies do not employ a uniform definition. Table B1 therefore summarizes the provenance and method used to determine $\alpha'$ for each data set. Where available, directly reported or reconstructed translational-stroke values were used; otherwise, representative values were estimated from published species-specific hovering kinematics. The resulting $\alpha'$ should therefore be regarded as a characteristic aerodynamic angle rather than a uniformly defined cycle-averaged quantity.

Table B1. Representative angle-of-attack values used in the hovering-insect analysis and their provenance and method of determination.

| $\alpha'$ source | data set | $\alpha'$ | determination |
|---|---|---|---|
| [18] | *Bombus terrestris* | 42.5° | directly reported / reconstructed from measured hovering kinematics |
| [19] | *Manduca sexta* | 35° | directly reported / reconstructed from measured hovering kinematics |
| [27] | *Coccinella septempunctata* | 45° | estimated from reported climbing-flight kinematics* |
| [28] | *Episyrphus balteatus* | 40° | estimated from hovering kinematics |
| [21] | *Eristalis tenax* | 33° | adopted from directly measured 3-d hovering kinematics |
| [22] | *Apis mellifera*, Wu & Sun data | 40° | estimated from reported hovering kinematics |
| [28] | *Bombus hortorum* | 25° | estimated from hovering kinematics |
| [22] | *Apis mellifera* | 55° | directly reported / reconstructed from measured hovering kinematics |
| [23] | *Bombus terrestris* | 28° | reported/model kinematics |
| [23] | *Manduca sexta* | 32° | reported/model kinematics |
| [21] | *Eristalis tenax*, DF1 | 33.8° | directly measured 3-d hovering kinematics |
| [21] | *Eristalis tenax*, DF2 | 35.5° | directly measured 3-d hovering kinematics |
| [21] | *Eristalis tenax*, DF3 | 34.2° | directly measured 3-d hovering kinematics |
| [24] | *Bombus ignitus*, individual 1 | 52° | directly measured hovering kinematics |
| [24] | *Bombus ignitus*, individual 2 | 46° | directly measured hovering kinematics |
| [24] | *Bombus ignitus*, individual 3 | 48° | directly measured hovering kinematics |
| [24] | *Bombus ignitus*, individual 4 | 47° | directly measured hovering kinematics |
| [25,29] | *Macroglossum stellatarum* | 40° | representative value from reported hovering kinematics |
| [26] | *Agrius convolvuli* | 47.5° | representative value from reported hovering kinematics |

*Hovering angle-of-attack data were unavailable and therefore climbing-flight kinematics were used.

## Appendix C: Minimum Aerodynamic-Power Condition

A first-order estimate of the aerodynamic power required for hovering can be obtained from $P_{\text{aero}} \approx \bar{F}_D \bar{V}$, where $\bar{F}_D$ is the mean aerodynamic drag and $\bar{V}$ is the cycle-averaged wingtip velocity. Extending the stationary-wing relationship $C_D = C_L \tan\alpha'$ observed with leading-edge perturbations [15], together with the hovering force balance $mg = \frac{1}{2}\rho \bar{V}^2 R c_m \bar{C}_L$, gives $\bar{F}_D = mg\tan\alpha'$ and hence (eliminating $\bar{V}$) gives:

$$P_{\text{aero}} \propto \tan\alpha' / \sqrt{\bar{C}_L}. \tag{7}$$

Substitution of the universal-$St_f$ loading relation from equation (4) allows the minimum-power angle to be determined analytically. Defining $B = (St_f \Lambda_s \Phi)^{-2}$, minimization of equation (7), gives

$$3(4B+1)\cos^4\alpha' - (24B+5)\cos^2\alpha' + 12B = 0, \tag{8}$$

and therefore

$$\alpha'_{\text{opt}} = \cos^{-1}\left[\left(\frac{24B+5-\sqrt{96B+25}}{6(4B+1)}\right)^{1/2}\right]. \tag{9}$$

Taking $St_f = 0.16$ and applying equation (9) to the insect geometries and stroke amplitudes in Table 1 gives $33^\circ \lesssim \alpha'_{\text{opt}} \lesssim 51^\circ$, with a mean of approximately $42^\circ$. This is substantially below the corresponding maximum-loading angles of approximately $57^\circ \lesssim \alpha'_{\max} \lesssim 64^\circ$. The result should be interpreted as a first-order minimum aerodynamic-power condition; rotational, inertial and three-dimensional contributions to flapping power are not included.

.